\documentstyle[12pt,epsf]{article}
\def\order{{\cal O}}

\newskip\zatskip \zatskip=0pt plus0pt minus0pt
\def\matth{\mathsurround=0pt}
\def\lsim{\mathrel{\mathpalette\atversim<}}
\def\gsim{\mathrel{\mathpalette\atversim>}}
\def\atversim#1#2{\lower0.7ex\vbox{\baselineskip\zatskip\lineskip\zatskip
  \lineskiplimit 0pt\ialign{$\matth#1\hfil##\hfil$\crcr#2\crcr\sim\crcr}}}

\begin{document}

\font\fortssbx=cmssbx10 scaled \magstep2
\hbox to \hsize{
\hskip.5in \raise.1in\hbox{\fortssbx University of Wisconsin - Madison}
\hfill$\vcenter{\hbox{\bf MAD/PH/785}
            \hbox{August 1993}}$ }

\vspace{.2in}

\begin{center}{\large\bf Astroparticle Physics with High Energy\\[.05in]
Neutrino Telescopes\footnotemark}\\[.2in]
Francis Halzen\\[.1in]
{\it Department of Physics, University of Wisconsin, Madison, WI 53706}
\end{center}

\footnotetext{Talk presented at the {\it Escuela Latino Americana de Fisica,
Mar del Plata (Argentina)}, at the {\it 14th International Workshop on Weak
Interactions, Seoul (Korea)}, and at the {\it Johns Hopkins Workshop on
Particles and the Universe, Budapest (Hungary)}, July 1993.}

\renewcommand{\LARGE}{\large}
\renewcommand{\Large}{\large}
\renewcommand{\thesection}{\arabic{section}.}

\vskip.2in

\begin{abstract}
The first optical modules of the Baikal high energy neutrino telescope have
recently been deployed. Commissioning of the AMANDA, DUMAND and NESTOR
detectors will follow soon. Before discussing the detectors we review the
arguments that pinpoint $0.1\rm~km^2$ as the natural scale of a neutrino
telescope. Though present detectors do not quite reach this goal, their
techniques, if successful, can be exploited to build km$^2$ detectors for a
cost not exceeding one hundred million dollars. Motivations for the
construction of km$^2$ deep underground detectors include

i) neutrino astronomy and the search for cosmic accelerators: we will focus
our discussion on recent claims that active galactic nuclei are the
accelerators of the highest energy cosmic rays. If this is true they are
inevitably high-flux sources of very energetic neutrinos;

ii) neutrino oscillations using the atmospheric neutrino beam: we emphasizing
the unique capability of surface neutrino telescopes, i.e. detectors
positioned at a depth of roughly $1\rm~km$, to detect neutrinos and muons of
similar energy. In a $\nu_{\mu}$ oscillation experiment one can therefore tag
the $\pi$ progenitor of the neutrino by detecting the muon produced in the
same decay. This eliminates the model dependence of the measurement inevitably
associated with the calculation of the primary cosmic ray flux. We will show
that planned surface neutrino telescopes probe the parameter space $\Delta m^2
\gsim 10^{-3}\rm\ eV^2$ and $\sin^2 2\theta \gsim 10^{-3}$ using this
technique. Recently underground experiments have given hints for neutrino
oscillations in this mass range.

iii) the search for neutrinos from the annihilation of dark matter
particles in our galaxy;

iv) the possibility to observe the thermal neutrino emission from supernovae
even though the nominal threshold of the detectors exceeds the neutrino energy
by several orders of magnitude.

v) to make the serendipitous discovery. No astronomical telescope,
detecting photons of any wavelength, has ever viewed sites in the Universe
shielded by more than a few hundred grams of matter.
\end{abstract}

\thispagestyle{empty}

\newpage

\section{Introduction}

With supernova 1987A underground detectors became very credible astronomical
telescopes\cite{Kamioka,IMB-2}. Their earlier mission was mainly particle
physics: proton decay, neutrino oscillations and the like. It was soon
realized, however, that the natural scale of a neutrino telescope is 1~km$^2$.
The only guaranteed source of high energy neutrinos I am aware of is the plane
of our own galaxy. This source is guaranteed by the very existence of high
energy cosmic rays. A flux of diffuse neutrinos from decay of charged pions is
produced when cosmic ray nuclei interact with interstellar gas. Even from the
direction of a dense target such as the galactic center, the ratio of
neutrinos and cosmic rays is less than $10^{-4}$ and a detection area of at
least 1~km$^2$ is required to observe the effect\cite{BER}.

Fortunately, astronomy, as well as the search for neutrino mass, for dark
matter, and the detection of supernova is possible with smaller detectors.

\section{The Gamma-Neutrino Connection}

Although observations of PeV ($10^{15}$~eV) and EeV ($10^{18}$~eV) gamma-rays
are controversial, cosmic rays of such energies do exist and their origin is
at present a mystery. The cosmic-ray spectrum can be understood, up to perhaps
1000 TeV, in terms of shock wave acceleration in galactic supernova
remnants\cite{VII}. Although the spectrum suddenly becomes steeper at
1000~TeV, a break usually referred  to as the ``knee\rlap,'' cosmic rays with
much higher energies are observed and cannot be accounted for by this
mechanism. This can be understood by simple dimensional analysis as the EMF in
the supernova shock is of the form
\begin{equation}
E = ZeBRc \,, \label{SN EMF}
\end{equation}
where $B$ and $R$ are the magnetic field and the radius of the shock. For a
proton Eq.~(\ref{SN EMF}) yields a maximum energy
\begin{equation}
E_{\rm max} = \left[10^5\,{\rm
TeV}\right]\left[B\over3\times10^{-6}\rm\,G\right]
\left[R\over50\,{\rm pc}\right]
\end{equation}
and therefore $E_{\rm max}$ is less than $10^5$~TeV for the typical values of
$B,R$ shown. The actual upper limit is much smaller than the value implied by
the dimensional argument.

Cosmic rays with energy in excess of $10^{20}$~eV have been observed. The
measured spectrum tells us that $10^{34}$ particles are accelerated to
1000~TeV energy every second. We do not know where or how. We do not know
whether they are protons or iron or something else. If their cosmic
accelerators exploit the $3 \mu$Gauss field of our galaxy they must be much
larger than supernova remnants in order to reach $10^{21}$~eV energies.
Eq.~(\ref{SN EMF}) indeed requires that their scale be of order 30~kpc and
this exceeds the dimensions of our galaxy. Although imaginative arguments
exists to avoid this impasse, an attractive alternative is to look for large
size accelerators outside the galaxy. Nearby active galactic nuclei (quasars,
blazars\dots) distant by order 100~Mpc are the obvious candidates. With
magnetic fields of tens of $\mu$Gauss over distances of kpc acceleration to
$10^{21}$~eV is possible; see Eq.~(\ref{SN EMF}).

One can visualize the AGN accelerator in a very economical way in the
Blanford-Zralek mechanism. The horizon of the black hole acts as a rotating
conductor immersed in an external magnetic field. By simple dimensional
analysis this creates a voltage drop
\begin{equation}
{\Delta V\over 10^{20}{\rm volts}} = {a\over M_{\rm BH}} \, {B\over 10^4{\rm
G}} \, {M_{\rm BH}\over 10^9 M_\odot} \,,
\end{equation}
corresponding to a luminosity
\begin{equation}
{{\cal L}\over10^{45}\rm erg\,s^{-1}} = \left(a\over M_{\rm BH}\right)^2
\left(B\over 10^4\rm G\right)^2 \left(M_{\rm BH}\over 10^9M_\odot\right)^2\,.
\end{equation}
Here $a$ is the angular momentum per unit mass of a black hole of mass $M_{\rm
BH}$.

All this was a theorist's pipe dream until recently the Whipple collaboration
reported the observation of TeV (10$^{12}$\,eV) photons from the giant
elliptical galaxy Markarian~421\cite{Mkr}. With a signal in excess of
6~standard deviations, this is the first convincing observation of TeV gamma
rays from outside our Galaxy. That a distant source such as Markarian~421 can
be observed at all implies that its luminosity exceeds that of galactic cosmic
accelerators such as the Crab, the only source observed by the same instrument
with comparable statistical significance, by close to 10 orders of magnitude.
More distant by a factor $10^5$, the instruments's solid angle for
Markarian~421 is reduced by $10^{-10}$ compared to the Crab. Nevertheless
the photon count at TeV energy is roughly the same for the two sources. The
Whipple observation implies a Mrk~421 photon luminosity in excess of $10^{43}$
ergs per second. It is interesting that these sources have their highest
luminosity above TeV energy, beyond the wavelengths of conventional astronomy.

Why Markarian 421? Whipple obviously zoomed in on the Compton Observatory
catalogue of active galaxies (AGN) known to emit GeV photons. Markarian, at a
distance of barely over 100~Mpc, is the closest blazar on the list. Stecker et
al.\cite{ir} recently pointed out that TeV gamma rays are efficiently
absorbed on infra-red starlight, anticipating that TeV astronomers will have a
hard time observing powerful quasars such as 3C279 at a redshift of 0.54.
Production of $e^+e^-$ pairs by TeV gamma rays interacting with IR background
photons is the origin of the absorption. The absorption is, however, minimal
for Mrk~421 with $z=0.03$, a distance close enough to see through the IR~fog.

This observation was not totally unanticipated. Many theorists\cite{HENA}
have identified blazars such as Mrk~421 as powerful cosmic accelerators
producing beams of very high energy photons and neutrinos. Acceleration of
particles is by shocks in the jets (or, possibly, in shocks in the accretion
flow onto the supermassive black hole which powers the galaxy) which
are a characteristic feature of these radio-loud active galaxies. Many
arguments have been given for the acceleration of protons as well as
electrons. Inevitably beams of gamma rays and neutrinos from the decay of
pions appear along the jets. The pions are photoproduced by accelerated
protons on the dense target of optical and UV~photons in the galaxy. The
latter are the product of synchrotron radiation by electrons accelerated along
with the protons. There are of course no neutrinos without proton
acceleration. The arguments that protons are indeed accelerated in AGN are
rather compelling. They provide a ``natural'' mechanism for i) the energy
transfer from the central engine over distances as large as 1 parsec, ii) the
heating of the dusty disc over distances of several hundred parsecs and iii)
the near-infrared cut-off of the synchrotron emission in the jet. Protons,
unlike electrons, efficiently transfer energy over large distances in the
presence of high magnetic fields.

Powerful AGN at distances of order 100 Mpc and with proton luminosities of
10$^{46}$\,erg/s or higher are therefore obvious candidates for the cosmic
accelerators of the highest energy cosmic rays, especially those with energy
in excess of 10$^{18}$~eV which are not confined by the magnetic field of our
galaxy. The neutrino flux from such accelerators can be calculated by energy
conservation
\begin{equation}
{\cal L}_p N\epsilon_{\rm eff} = 4\pi d^2 \int dE [E\, dN_\nu / dE] \;,
\end{equation}
where $N_\nu$ is the flux at Earth, $d$ the average distance to the sources,
$N$ the number of sources and $\epsilon_{\rm eff}$ the efficiency for protons
to produce pions. We assume order 1 neutrino per interacting proton. This
yields
\begin{equation}
E {dN_\nu\over dE} = {N \epsilon_{\rm eff}\over 4\pi}
{1.4\times 10^{-8} \over E\,\rm(TeV)} \,\rm cm^{-2}\, s^{-1}\, sr^{-1} \;.
\end{equation}
for ${\cal L}_p=10^{46}$\,erg/s and $d=100$~Mpc. With $\epsilon_{\rm eff}$ of
order
$10^{-1}$ to $10^{-2}$ and $N$ in the range 1 to 100, we obtain neutrino fluxes
which are well within reach of $0.1\rm~km^2$ neutrino telescopes. For
$N\epsilon_{\rm eff}=1$ we obtain a diffuse neutrino flux which is a factor 2
larger than the one predicted by a detailed model worked out by Biermann and
collaborators\cite{bier}. It successfully accomodates the observed spectrum of
cosmic rays with energy in the EeV range. We predict 610 up-coming muon events
per year in a $10^5\rm\,m^2$ detector. The calculation is straightforward and
follows reference\cite{muon}. Neutrino absorption in the Earth has been
included. The relative large contribution of very high energy neutrinos to the
signal is otherwise overestimated.

We now turn our attention to individual sources. We take Mrk~421 as an example
and normalize all calculations to the observed high energy photon flux
\begin{equation}
\int_{1/2\,\rm TeV} dE { dN_\gamma\over dE } =
1.5\times 10^{-11} \, \rm cm^{-2}\, s^{-1} \;. \label{norm}
\end{equation}
We will assume that $N_\gamma$ has a power spectrum with spectral index
$\gamma$.
We work with the blazar jet model of Biermann et al.\cite{HENA} which
accommodates fluxes of active galaxies at all wavelengths and whose
predictions we have parametrized in a convenient form. The maximum particle
energies are given by
\begin{equation}
E_{p\,\rm max} \simeq 10 E_{\gamma\,\rm max} =
2\times 10^{20}\,{\rm eV} \left[\Gamma\over 10\right]
\left[ 1\,\rm Gauss\over B \right]^{1/2} \;,
\end{equation}
where $\Gamma$ is the Doppler shift to the comoving frame, {\it i.e.}\ from the
frame where acceleration takes place to our Earth frame. One has to realize
that for some blazars the jet is beaming particles in our direction. The
optical depth of the source, {\it i.e.}\ the jet, is given by
\begin{equation}
        \tau_{\rm optical} = 2 \left[ B\over 1 \,\rm Gauss\right]^{1/2}
        \left[ E_{\gamma}\over \rm 1\, TeV\right] \;. \label{tau_opt}
\end{equation}
For a 1 G field the optical depth of the source is unity for the 0.5~TeV
photons observed by Whipple. Although the value of the $B$-field in the jet is
a guess which ranges from $10^{-4}$ to $10^4$~G, $10^3$~G is the
typical assumed value.

We compute the gamma ray flux inside the source by correcting the observed
flux (\ref{norm}) for absorption in the jet. The optical depth is given by
(\ref{tau_opt}). The answer depends critically on the magnitude of the
$B$-field which is understandable because photon energy loss inside the source
is primarily on the magnetic field in the jet. We subsequently estimate the
neutrino flux assuming 1 neutrino per gamma ray, which should approximately
hold for pion decay, i.e. $N_\nu =  \left[ N_\gamma \right]_{\rm before\
absorption}$.

Our results are shown in Table 1 for a range of assumptions for $B$ and the
spectral index $\gamma$. We conclude that, even with the most pessimistic
assumptions, Mrk~421 should produce a handful of upcoming muon events per year
in our generic $10^5\rm\,m^2$ detector. A 1~km$^2$ detector is necessary for
guaranteed detection. Notice, however, that for most of the $B,\gamma$
parameter space the predicted event rates are much larger. For $B$-fields of
100~G and more the predicted rate is so large that even the smallest
underground detector should already have detected Mrk~421. This leads to some
interesting considerations. Because the source is positioned in the northern
hemisphere, only a southern neutrino telescope is sensitive. This excludes all
underground facilities except for the small KGF experiment which has
relatively poor angular resolution. Soon AMANDA, under construction at the
South Pole, will have an uninterrupted view of Mrk~421.

\begin{center}
\parbox{5.5in}{\small Table 1: Number of upcoming muons ($N$) per $
10^5\rm\,m^2$ per y for the different scenarios. ${\cal L_\gamma}$ is in
$10^{43}$\,erg/s}
$\begin{array}{|c|c|c|c|c|c|}
\hline
\mbox{B (Gauss)} & \gamma & E_{p,\rm max}
& E_\gamma \; \mbox{for} \; \tau_{\rm opt}=1 & {\cal L_\gamma} & N \\
\hline
        & 1   &                     &                   & 30 & 2 \\
10^{-4} & 0.8 & 2 \times 10^{22}\rm eV & 50 \;  \mbox{TeV} & 500 & 11 \\
        & 0.4 &                     &                   & 10^6 & 450 \\
\hline
1       & 1   & 2 \times 10^{20}\rm eV & 500 \; \mbox{GeV} & 200 & 13 \\
\hline
10      & 1   & 6 \times 10^{19}\rm eV & 150 \; \mbox{GeV} & 3000 & 270 \\
\hline
10^{4}  & 1   & 2 \times 10^{18}\rm eV & 5 \;   \mbox{GeV} & -
& \mbox{{\small unreasonable}} \\
\hline
\end{array}
$
\end{center}

It is important to point out that while the assumption that $\gamma=1$, {\it
i.e.}\ a normal $E^{-2}$ energy spectrum, is usually optimistic, in this case
it is not. In the blazar model where acceleration is by shocks in the jet
rather than in the accretion disc, $\gamma$ is actually zero near $E_{\rm
max}$. The reason is interesting. The photon target on which the protons
photoproduce pions has itself an $E^{-2}$ spectrum. As protons are accelerated
they encounter an increasing number of photons with sufficient energy to
photoproduce pions. This is the origin of the very flat spectrum. The neutrino
output of a blazar can be conveniently parametrized as\cite{HENA}
\begin{equation}
E {dN_\nu\over dE} = 1.6\times 10^{-17}\,{\rm cm^{-2} s^{-1}}
\left[{\cal L_\gamma}\over 10^{48}\rm erg/s\right] \left[0.54\over z\right]^2
\left[10\over \Gamma\right] \left[B\over 1\,\rm Gauss\right]^{1/2} \;.
\end{equation}
The formula is scaled to 3C279. Predictions for neutrino event rates based on
this formula are shown in Fig.~1 for Mrk 421 and for the total luminosities
$10^{46}$ (solid), $10^{47}$ (dotted) and $10^{48}$ erg/s (dashed), upper
(lower)  curves are without (with) earth absorption. The event
rates are similar to those previously estimated on the basis of the Whipple
observation.

\begin{center}
\epsfxsize=4.25in\hspace{0in}\epsffile{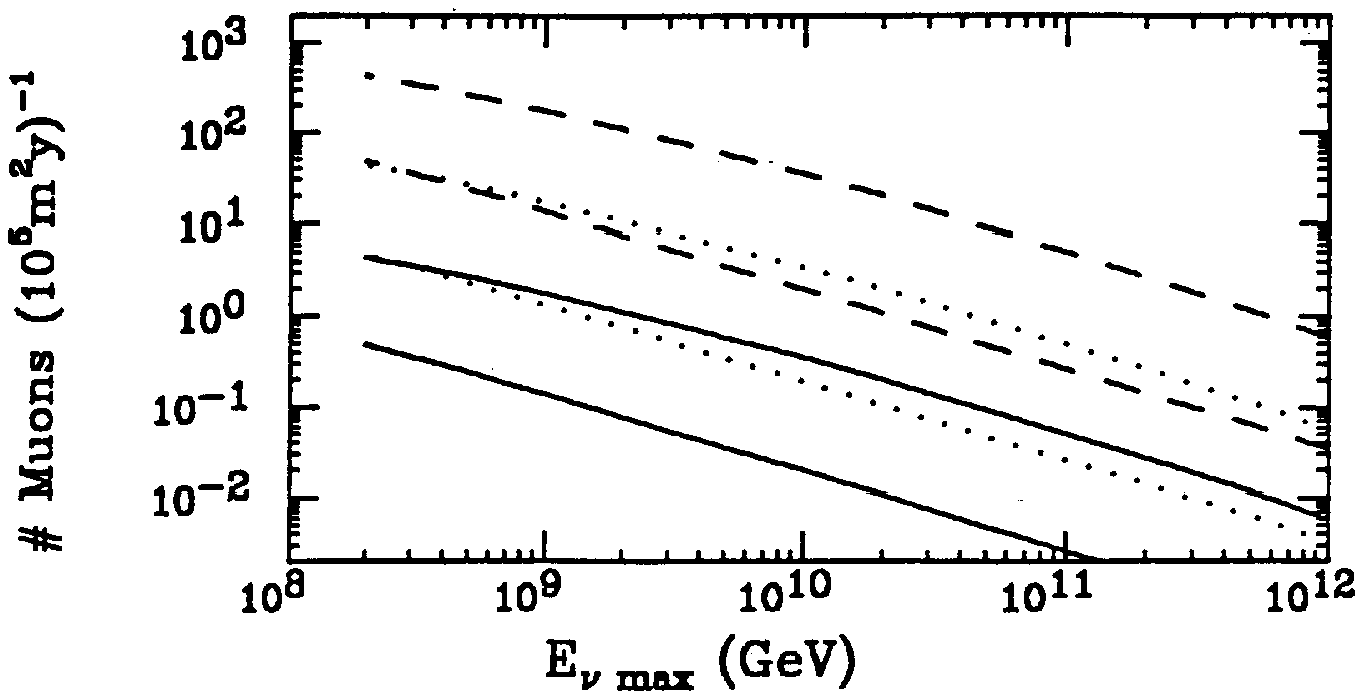}

\parbox{5.5in}{\small Fig.~1: Predictions for neutrino event rates  for Mrk~421
and for the total luminosities $10^{46}$ (solid), $10^{47}$ (dotted) and
$10^{48}$ erg/s (dashed), upper (lower)  curves are without (with) earth
absorption.}
\end{center}

One should appreciate that weakly interacting neutrinos will make their way to
our detectors unattenuated by ambient matter in the source or by IR
light. So, while high energy photons are absorbed on intergalactic IR photons
for AGNs much further than Mrk~421, neutrinos are not and sources should be
detected with no counterpart in high energy photons. The known AGN's form a
cast of thousands and theorists have estimated that their collective flux will
actually dominate the atmospheric flux for PeV neutrino energies.

\section{Neutrino Oscillations in the Atmospheric Neutrino Beam}

Data from the {\sc KAMIOKANDE}\cite{kamos} and IMB\cite{imbos} collaborations
indicate
a deficit in the measured ratio of $\nu_{\mu} / \nu_e$ atmospheric neutrino
events when compared with a calculation of their relative flux in the
atmospheric cosmic ray beam. There has been considerable interest in
experiments capable of verifying this result, e.g.\ long-baseline experiments
observing an accelerator neutrino beam in an underground detector located
elsewhere\cite{P}. The advantages of such experiments are important:
i) a long baseline is obtained, e.g. up to 6000~km between Fermilab and the
detector\cite{lp} in Hawaii, and ii) the properties of the accelerator
neutrino beam are known (or can be measured on site), thus eliminating the
model dependence associated with the calculation of the primary cosmic ray
fluxes in the {\sc KAMIOKANDE} and IMB experiments. Using this technique it is
anticipated that the parameter space $\Delta m^{2} \gsim 10^{-3}$ and $\sin^2
2\theta \gsim 10^{-3}\rm\ eV^2$ can be probed\cite{P}. A new generation of
surface neutrino telescopes can achieve the same goals by measuring the ratio
of up- and down-going muons. The fact that surface detectors, unlike
underground experiments, can measure up-going $\nu_{\mu}$'s and down-going
$\mu$'s of similar energy will be crucial.

The argument can be best introduced as follows. The {\sc KAMIOKANDE}
experiment typically observes 500~MeV neutrinos which are, on average, the
decay products of 20~GeV pions. Instead of calculating the 20~GeV pion flux,
on which the oscillation measurement depends, we could measure it by tagging
the muon in the $\pi \rightarrow \mu \nu_{\mu}$ decay along with the
oscillating neutrino. Such GeV-energy muons, produced high in the atmosphere,
do not reach sea level. In order to count them one could fly the detector in a
balloon. Doing this experiment has, in fact, been advocated\cite{perk}
although the use of a different detector is quite inevitable. This
straightforward idea does not really work as $\nu_{\mu}$ are also produced by
$\mu \rightarrow e \nu_e \nu_{\mu}$ decay and this muddles the analysis.

Our main point is that such measurement can be performed in surface neutrino
telescopes\cite{don}. The ratio of muon energy to the neutrino-induced muon
energy from the same pion parent is somewhat less than a factor 10. The
threshold for neutrino detection in a detector such as AMANDA is between
2--50~GeV and the tagging muons therefore have energies in the 200~GeV range
by pion decay kinematics. This is precisely the threshold by energy-loss for
down-going muon in this instrument positioned under 1~km of polar ice. The
parent pion flux of the oscillating neutrinos can therefore be tagged in the
relevant energy range by measuring the down-going muon flux. Obviously the
atmospheric parent pion flux of such muons is the same above the detector as
on the other side of Earth where the detected $\nu_{\mu}$ originated. Also, at
these energies muons do not decay and there are no other significant sources
of cosmic ray muons. The down-going muons therefore determine the atmospheric
pion and $\nu_{\mu}$ flux. The necessity of a calculation has thus been
eliminated. In reality, of course, some secondary dependence on Monte Carlo
calculations will remain, {\it e.g.} to account for the fact that the
detectors are unlikely to have perfectly symmetric up/down acceptance. The
primary handicap in atmospheric neutrino experiments, that the flux of the
primary beam is less well understood than in accelerator neutrino experiments,
is essentially eliminated.

The key variables in evaluating the capabilities of the experiment are the
length of the baseline, the muon energy threshold and the minimum measurable
oscillation probability $\epsilon$. The last variable parametrizes the
inherent systematical errors of the measurement\cite{P}. It represents the
``quality'' of the instrument.  We will vary $\epsilon$ between an optimistic
1\% and a rather pessimistic 10\%.

For illustration we study the sensitivity or $\epsilon$ plots in the $\nu_{\mu}
\leftrightarrow \nu_{\tau}$ case. Fig.~2(a) shows the region of $\Delta m^2_0$
and $\sin^2 2 \theta_0$ space that can be probed by an experiment sensitive to
${\cal P}_{ab} > \epsilon$, for $\epsilon = 1 \%$, $3 \%$ and $10 \%$.
Previous discussions of long-baseline neutrino oscillation experiments have
usually stopped here, after arguing that systematic errors were reasonably
under control. It is, however, essential to realize that the statistical
errors involved will be quite important in these experiments. This is true
whether one uses the accelerator or cosmic ray beam. We emphasize that it is
critical to introduce criteria for observation of oscillation based on
statistical errors. We illustrate this point by calculating the excluded
regions in $\Delta m^2_0$ and $\sin^2 2\theta_0$ for 4$\sigma$ deviations from
the expected muon detection rate after 1~year of operation; see Fig.~2(b).

\vspace{-.3in}

\begin{center}
\epsfxsize=3.5in\hspace{0in}\epsffile{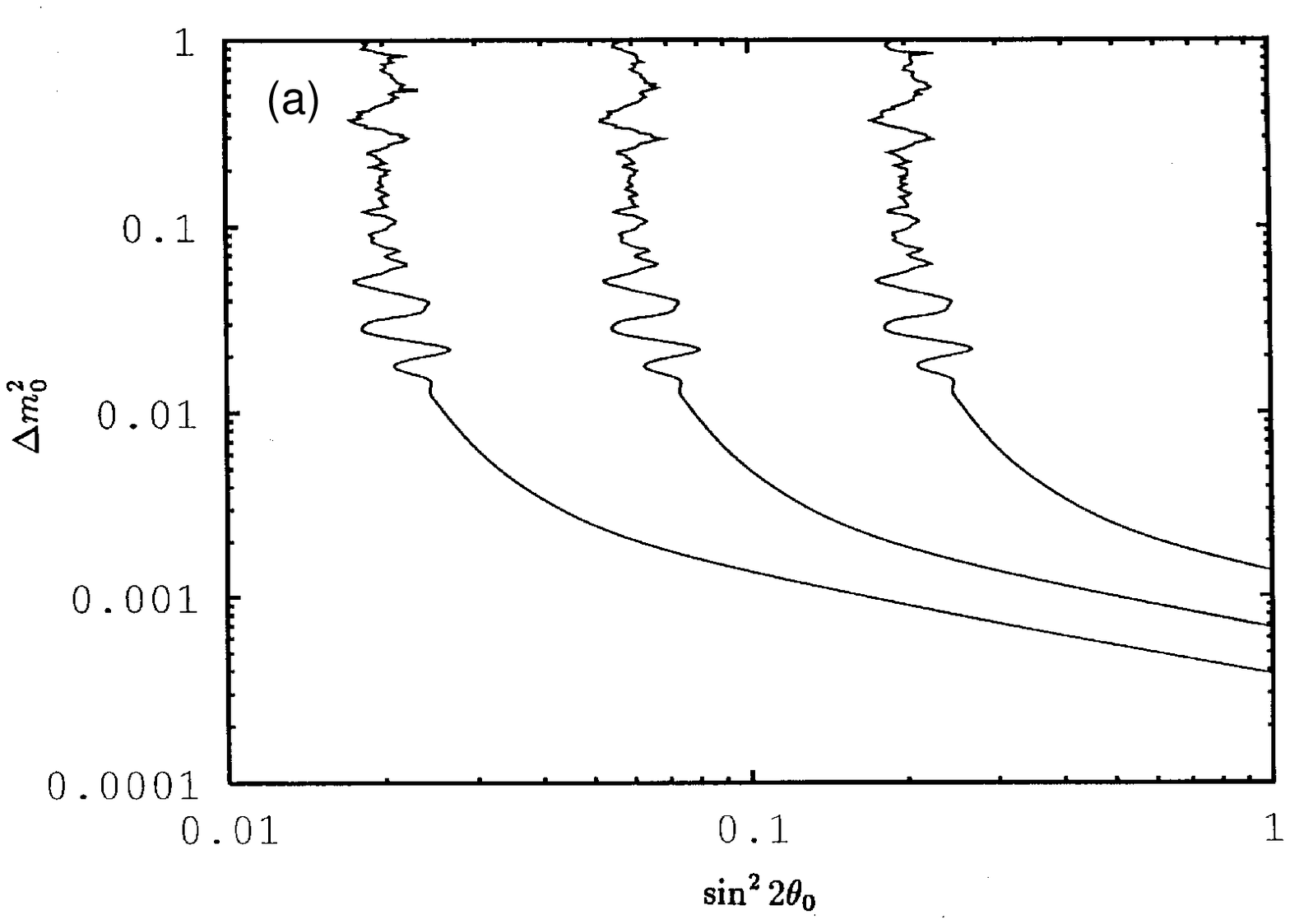}

\parbox{5.5in}{\small Fig.~2(a): The search regions (in $\Delta m_0^2$ vs.
$\sin^2 2 \theta_0$) based on a (left to right) 1\%, a 3\% and a 10\%
measurement of the $\nu_\mu \leftrightarrow \nu_\tau$ oscillation probability.}

\smallskip

\epsfxsize=3.5in\hspace{0in}\epsffile{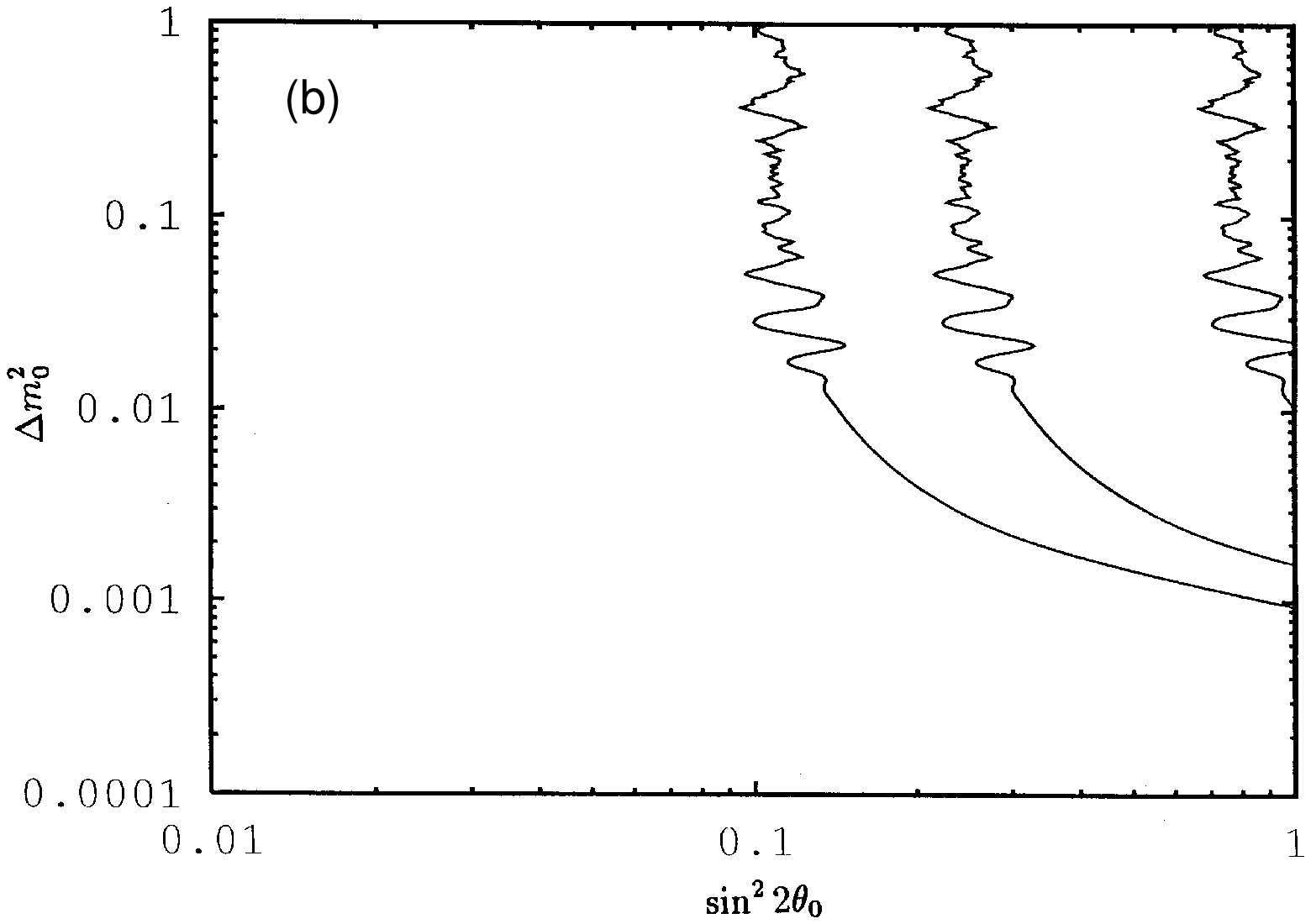}

\parbox{5.5in}{\small Fig.~2(b): The $4 \sigma$ measurement search regions for
surface neutrino detector with area of (left to right) $10^5\rm\,m^2$, $2
\times10^4\rm\,m^2$ and $2 \times 10^3\rm\,m^2$.  This figure is for the case
of $\nu_\mu\leftrightarrow \nu_\tau$ oscillation.}
\end{center}

The sensitivity of AMANDA is in the end similar to that of long-baseline
experiments using accelerator neutrinos. We will use for comparison the {\sc
Dumand}\cite{lp} detection of a Fermilab neutrino beam.  Calculating the
detected event rates following Refs.\cite{lp} and the Fermilab
neutrino beam spectrum given in Ref.\cite{P}, we show the $4 \sigma$ reach in
$\Delta m^2$, $\sin^2 2\theta$ of the long-baseline experiment in Fig.~3 for
$\nu_{\mu} \leftrightarrow \nu_{\tau}$ oscillations. We again assumed one year
of running. With a baseline of 6000~km and a threshold of 20~GeV the
observable oscillations are essentially identical to those of a generic
surface neutrino telescope of similar area, {\it i.e.} 20000~km$^2$ for the
example shown. The plot shows residual $\sin^2 (1.27 \Delta m^2_0 L/K)$
oscillations behavior, which is absent in the surface detector plots due to
the broader (in energy) atmospheric neutrino flux and the varying neutrino
paths $L$ through the Earth. The net result is still that the two different
types of experiments are able to place comparable limits on $\Delta m_0^2$ and
$\sin^2 2 \theta_0$, though they are able to exclude somewhat different
regions of parameter space.

\begin{center}
\epsfxsize=4in\hspace{0in}\epsffile{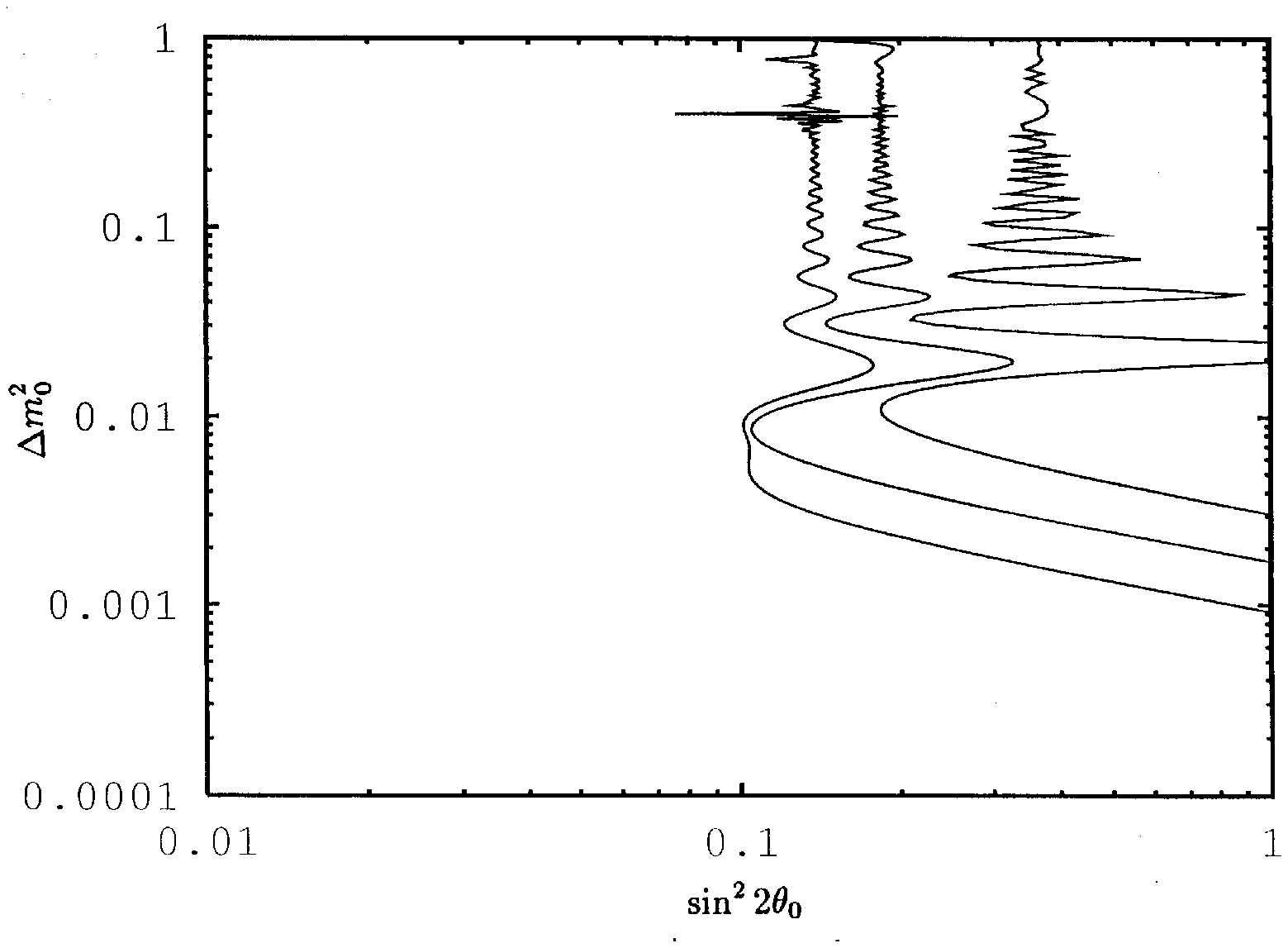}

\parbox{5.5in}{\small Fig.~3: The $4 \sigma$ measurement search regions for
long-baseline experiment using the Fermilab neutrino beam with muon thresholds
of (left to right) 10 GeV, 20 GeV and 40 GeV.  This figure is for the case of
$\nu_\mu \leftrightarrow \nu_\tau$ oscillation.}
\end{center}

\section{Neutrino Telescopes as Dark Matter Detectors}

There is compelling evidence for the presence of dark matter in our galaxy.
Some dark matter particle candidates, e.g.\ WIMPS, annihilate into photons,
electrons and protons thus providing an indirect signature for their
existence. They also annihilate into neutrinos and neutrino telescopes can
therefore also be used to search for dark matter.

If supersymmetry is Nature's extension of the Standard Model it must produce
new phenomena on the scale of several TeV or below. A very attractive feature
of supersymmetry is that it provides cosmology with a natural dark matter
candidate\cite{Ellis}. The supersymmetric partners of the photon, $Z^0$, and
the two Higgs particles form four neutral states, the lightest of which is the
stable neutralino.  If this lightest supersymmetric particle has an
appreciable relic abundance it may also be responsible for the dark matter
known to exist in the galactic halo. Neutralinos in the halo will scatter off
elements in the Sun and become gravitationally trapped.  The trapped
neutralinos will annihilate, producing high-energy neutrinos which may be
detected on Earth. The KAMIOKANDE\cite{Kam} and Irvine-Michigan-Brookhaven
(IMB)\cite{IMB} collaborations have already demonstrated that this indirect
neutrino signature provides us with a powerful tool for searching for dark
matter.  They extended the limits on neutralino mass into the 30--80 GeV
window by using this technique.

If neutralinos have masses greater than a few TeV, they ``overclose'' the
Universe.  Supersymmetry has therefore been framed inside a well defined
GeV--TeV mass window. Here we ask the question, ``What size telescope is
required to search this range?'' We will conclude that neutrino detectors of
order 1 km$^2$ are required, although clearly progress is possible with any
experiment larger than KAMIOKANDE\cite{hks}.

The parameter space of supersymmetric models is complicated, even in the
so-called minimal supersymmetric standard model (MSSM). The model is specified
by the top mass $m_t$ and five parameters: two unphysical masses $M_2$ and
$\mu$, the ratio of the Higgs vacuum expectation values $\tan\beta=v_2/v_1$ ,
the mass of the lightest Higgs $M_{H_2}$, and the squark masses $M_{\tilde
q}$. Some of these parameters are constrained by accelerator searches and by
cosmological considerations. The ratio of the Higgs vacuum expectation values
is bracketed by the values ($1,m_t/m_b$), where $m_b$ is the bottom-quark
mass. Radiative corrections tend to drive the ratio above unity, and the upper
bound results from constraints on electroweak symmetry breaking in
supergravity models\cite{Giudice}. Unsuccessful accelerator searches have
pushed the lightest Higgs mass above 50 GeV\cite{LEP}. Finally, naturalness
and cosmology favor values of $M_2$ and $\mu$ less than or of the order of 10
TeV; some would argue much less. This still leaves us with a large parameter
space to search.  In order to perform a manageable analysis of the problem, we
choose $M_2$ and $\mu$ as our independent parameters, and fix all other
parameters to reasonable values.  We set $m_t=120$ GeV, $\tan\beta=2$,
$M_{H_2}$=50 GeV and, the squark masses to be infinite (which minimizes
interactions involving quarks and leptons, thereby maximizing the relic
abundance but minimizing capture rates). The dependence on these parameters is
discussed in reference \cite{hks}.

Once these parameters have been specified, standard Big-Bang cosmology can be
used to determine the relic abundances of neutralinos in the Universe.
Some of the parameter space can be readily eliminated since the corresponding
models result in an unacceptably large relic density, i.e.\
$\Omega_\chi h^2 \gsim
1$ where $\Omega_\chi$ is the fraction of critical density
contributed by neutralinos and $h$ is the Hubble parameter in
units of 100 km sec$^{-1}$ Mpc$^{-1}$.
On the other hand, since the fraction of critical density
contributed by halos is $\Omega_{\rm halo}\sim\order(0.1)$ and
$h$ may be $\sim0.5$, we say that if $\Omega_\chi h^2 \lsim 0.02$
the neutralino
relic abundance is cosmologically consistent but too small to
account for the halo dark matter. In models
where $\Omega_\chi\gsim \Omega_{\rm halo}$ it is reasonable to
assume that the galactic halo dark matter {\it is} made up of
neutralinos and that the local mass density in neutralinos is
0.4 GeV cm$^{-3}$.  The shaded regions in
Fig.~4 are thus excluded as dark-matter candidates; i.e. the
models do not satisfy $0.02 \lsim
 \Omega_\chi h^2 \lsim 1$.  The dependence of the relic density on
the squark mass is important: for a detailed discussion see
reference\cite{hks}.

\begin{center}
\epsfxsize=4in\hspace{0in}\epsffile{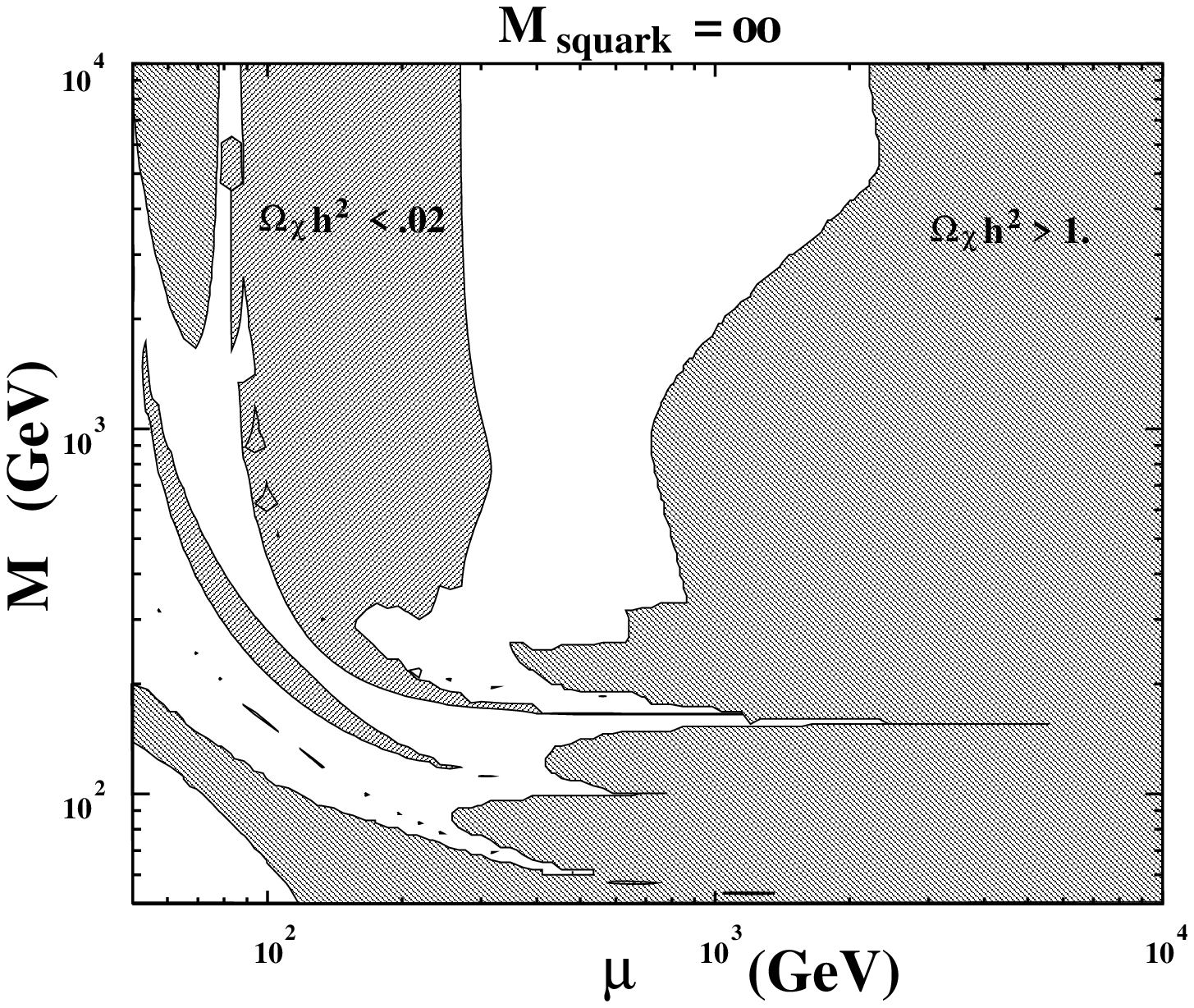}

\medskip

\parbox{5.5in}{\small Fig.~4: Regions in the $M_2$-$\mu$ plane, with $M_{\tilde
q}=\infty$, which are ruled out by cosmological considerations.  Upward
diagonal regions are areas where $\Omega_\chi h^2 >1$.  Downward diagonal
regions are areas where $\Omega_\chi h^2 < 0.02.$}
\end{center}

Our main conclusions follow from Fig.~5(a) which shows the
detector area required to observe one neutrino event per year. This
quantity is shown as a function of $M_2$ and $\mu$. Various annihilation
thresholds are clearly visible.  Most noticeable are the thresholds
associated with the $W$ and $Z$ masses near 100~GeV. The graphs
confirm that a detector of km$^2$ scale is required to study the full
neutralino mass range. The scatter plot of Fig.~5(b) shows the
information in Fig.~5(a) in a slightly different way.  Each dot gives
the event rate for a point in the $M_2$-$\mu$ plane in which the
neutralino is a good dark-matter candidate ($0.02\lsim\Omega_\chi h^2
\lsim 1$).  Note that the density of points is proportional to the area
in the $M_2$-$\mu$ plane. The figure clearly shows that, except for
some very special parameters, the discovery of supersymmetric dark
matter is within reach of any detector exceeding $10^5$ m$^2$ area; a more
realistic evaluation of the sensitivity of experiments can be found in
reference\cite{hks}. It is expected that the Earth will produce a signal
similar in
magnitude to the one calculated from the Sun. For the high end of the GeV--TeV
mass range investigated here, the signals from the Sun dominate by a factor of
roughly five\cite{Gould}. This result is however reversed for the lower
neutralino masses. For low-mass neutralinos, signals from the Earth will
significantly increase sensitivity.

\begin{center}
\epsfxsize=5in\hspace{0in}\epsffile{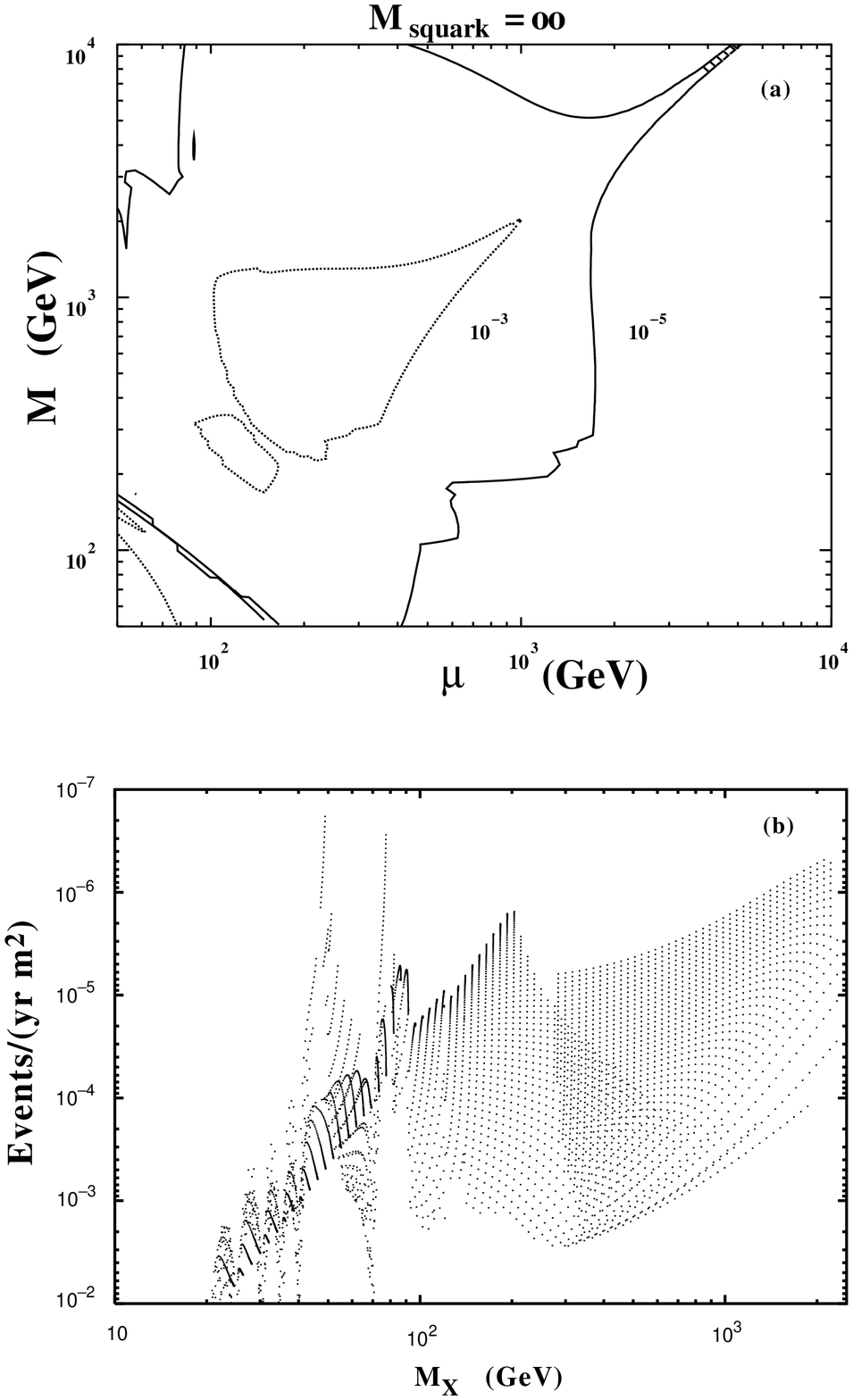}

\medskip

\parbox{5.5in}{\small Fig.~5: Events m$^{-2}$ yr$^{-1}$.  Figure (a) shows
contours of constant
detection rate in the $M_2$-$\mu$ plane for
$M_{\tilde q}=\infty$.  Figure~(b) shows the same information as
figure~(a) but in a slightly different way.
Each dot gives the event rate for a point in the $M_2$-$\mu$
plane in which the neutralino is a good dark-matter candidate
($0.02\lsim\Omega_\chi h^2 \lsim 1$).  Note that the density of points is
proportional to the area in the $M_2$-$\mu$ plane.
We fix $m_t=120$ GeV,
$\tan\beta=2$, $M_{H_2}=50$ GeV, and $M_{\tilde q}=\infty$.}
\end{center}

We conclude that a neutrino telescope of $10^5$~m$^2$ or more is clearly a
superb instrument to search for supersymmetric dark matter and offers great
promise for discovery should the galactic halo be composed of neutralinos. Its
failure to observe dark matter particles would force supersymmetrists to
fine-tune their models into small regions of parameter space. Progress in the
search for neutralinos can, however, be made with any detector larger than
KAMIOKANDE.

\section{Supernova Detection with High Energy Neutrino\hfill\break
 Telescopes?}

The neutrino events detected in the KAMIOKANDE\cite{Kamioka}, IMB\cite{IMB-2}
and, arguably, the Baksan\cite{Baksan} and LSD\cite{LSD} detectors prior to
the optical discovery of supernova 1987A represented a most remarkable birth
of neutrino astronomy. Despite observation by four experiments, the data has
left us with some lingering doubts. Most prominent is our inability to
understand the time of the Mont Blanc neutrino burst\cite{LSD} and the
directionality of the IMB events\cite{IMBang}. The success of the theory in
anticipating the features of the high statistics KAMIOKANDE and IMB data leads
us to suspect that these problems are instrumental rather than real. The
history of 1987A nevertheless underscores the importance of collecting as much
information as possible when presented with the rare opportunity of observing
the next nearby supernova. The high energy neutrino telescopes, discussed
here, can contribute information, even though their nominal neutrino threshold
are far too high.

First generation high energy neutrino telescopes consist of approximately 200
optical modules (OM) deployed in deep, clear water or ice shielded from cosmic
rays\cite{FHVenice}. Coincident signals between the OMs detect the \v Cerenkov
light of muons with energy in excess of a few GeV. Electromagnetic
showers initiated by very high energy electron neutrinos are also efficiently
detected. The idea has been debated for some time whether these instruments
have the capability to detect the MeV neutrinos from a supernova. The
production of copious numbers of positrons of tens of MeV energy in the
interaction of $\bar\nu_e$ with hydrogen, will suddenly yield signals in all
OMs for the 10 seconds duration of the burst. Clearly such a signal, no matter
how weak, will become statistically significant for a sufficient number of
OMs. We recently performed a complete simulation of the signal and its
detection and concluded that the 200 OMs of detectors such as DUMAND and
AMANDA are sufficient to establish the occurrence of a neutrino burst in
coincidence with the optical display of a supernova. A 1987A-type supernova in
the center of the galaxy will generate almost $10^4$ events in the AMANDA
detector resulting into an observation in excess of 5~$\sigma$. We also showed
that the same detectors can actually serve as a supernova watch, i.e.\ a
``fake" signal occurs less than once a century, by increasing the number of
OMs by a factor of three. This is much less than the roughly 7000~OMs which
are projected for a next-generation detector\cite{FHVenice}.

In the end the observations remarkably confirmed the established ideas
for the supernova mechanisms\cite{mechanism}. Most of the energy is liberated
after deleptonization in a burst lasting about ten seconds. Roughly equal
energies are carried by each neutrino species. The time scale corresponds to
the thermalization of the neutrinosphere and its diffusion within the dense
core. Since the ${\bar{\nu}}_e$ cross-section for the inverse beta decay
reaction on protons exceeds the characteristic cross sections for the other
neutrino flavors, ${\bar{\nu}}_e$ events dominate by a large factor after
including detection efficiency. In this reaction free protons absorb the
antineutrino to produce a neutron and a positron which is approximately
isotropically emitted with an energy close to that of the initial neutrino.
For the purpose of illustration we use typical parameters, derived from SN1987
observations, which are consistent with those previously estimated in
supernova models. From the energy distributions of the observed events the
average temperature of the neutrino sphere in SN1987A was deduced to be
4.0~MeV\cite{mechanism}.

It is straightforward to make a back-of-the-envelope derivation of the
distance over which optical modules in high energy neutrino telescopes can
detect supernova neutrinos. After convoluting the 4~MeV thermal Fermi
distribution of the neutrinos with a detection cross section rising with the
square of the neutrino energy, one obtains an event distribution peaked in the
vicinity of 20~MeV. The tracklength of a 20~MeV positron is roughly 10
centimeters and therefore over 3000 \v Cerenkov photons are produced. This
number combined with a typical quantum efficiency of $25\%$ leaves 800
detected photons in each event. It is easy to estimate either by analytic
calculation or by a detailed Monte Carlo simulation\cite{hjz}. For OMs such as
those used in the AMANDA detector with collecting area $A_{M}=0.028$~m$^2$ and
for an attenuation length $R_{\rm att}=25$~m typical of ice\cite{iceatt}, one
obtains $V_{\rm eff} \sim 130$~m$^3$ per optical module for detecting the
Cherenkov radiation from neutrino-induced electrons. This result can be used
to rescale SN1987 observations to a supernova at a distance $d_{kpc}$. From 11
events observed in 2.14~kton KAMIOKANDE detector we predict:
\begin{equation}
N_{Events} \sim 11~N_{M}~\left[{\rho~V_{\rm eff} \over 2.14\rm~kton} \right]
\left[ {52\rm~kpc \over d_{\rm kpc}} \right]^2
\end{equation}
for a detector with $N_{M}$ optical modules. Here $\rho$ is the density of ice.
For a 130~m$^3$ effective volume of each of the 200 OMs we obtain 5300 events.
A detailed simulation of the full detector leads to an event rate which is
50\% larger.

We next must require a meaningful detection of this signal in the presence of
the continuous background counting rate of all phototubes. Over the 10~s
duration of the delayed neutrino burst from a supernova, the rms fluctuations
of the combined noise from all the OMs is:
\begin{equation}
\sigma_{1p.e.} = {\sqrt {10~\nu_{1p.e.}~N_{M}}}
\end{equation}
where the background counting rate in each module at the 1 photoelectron level
is represented by $\nu_{1p.e.}$. The probability that the noise in the OMs
fakes a supernova signal can be estimated assuming Poisson statistics. The
expected rate of supernova explosions in our galaxy is about
$2\times10^{-2}$~y$^{-1}$. If the detector is to perform a supernova
watch we must require that the frequency of fake signals is well below this
rate. The signal should therefore exceed $n_{\sigma} \ge 6$ which
corresponds to a probability of $9.9\times10^{-10}$. The corresponding number
of 10~seconds intervals indeed exceeds a century. Clearly the requirement can
be relaxed if we just demand that the detector can make a measurement in the
presence of independent confirmation. For an average noise rate of 1~kHz, a
typical value for the OMs in AMANDA, the rms fluctuation of the 20 million hits
expected in an interval of ten seconds is 1400. This implies that detection of
a galactic supernova is near the 4~$\sigma$ level for the 200 module
configuration, while detection should not represent a problem for the next
generation detector which consists of 7,000~OMs. Since the signal in the
present detector is marginal, it is necessary to do a more realistic
calculation of the event rate. We will conclude that our rough estimate is
somewhat conservative.

Background noise in the modules clearly plays a critical role so that low
noise environments such as ice have an intrinsic advantage. Furthermore, the
noise is expected to be reduced drastically at the 2 photoelectron level. This
will unfortunately also imply a reduction in effective volume for event
detection as we will see further on. Signal to noise is proportional to the
ratio $V_{\rm eff}/{\sqrt{\nu}}$. Obviously increased attenuation length in the
medium and larger effective area of the OM results in an enhanced effective
volume. Considering parameters appropriate for DUMAND, an attenuation length
of 40~m in water and OMs with double diameter, we expect a factor 10 increase
in effective volume per optical module. This should readily compensate for a
noise rate higher by a factor 100. We therefore expect DUMAND and AMANDA to
have comparable sensitivity as supernova detectors. Detailed calculations can
be found in reference \cite{hjz}.

\section{Kilometer Scale Detectors?}

It should by now be clear that a high-energy neutrino telescope is a
multi-purpose instrument which can make contributions to astronomy,
astrophysics and particle physics.  Because photons, whatever their
wavelength, are absorbed by a few hundred grams of matter, high-energy
neutrinos provide us with a first opportunity to do a tomographic study of the
Universe. Therefore the hope, and greatest probability based upon the history
of forays into new wavelength regions, is to discover unanticipated phenomena.
 It is nevertheless important to simulate the performance of a future
telescope in all manner of more mundane physics circumstances in order to
determine its natural size. It is intriguing that our previous estimates all
point to the necessity of building 1~km$^3$ detectors\cite{LEA,HAL}. We
close with a discussion of the possibility of building a 1~km scale neutrino
detector based on the experience gained in designing the instruments now under
construction, specifically AMANDA, Baikal, DUMAND and NESTOR which we will
briefly review\cite{LEA92a,HEN}. One can confidently predict that such a
telescope can be constructed at a reasonable cost, e.g.\ a cost similar to
Superkamiokande\cite{SUZ}, to which it is complimentary in the sense that its
volume is over two orders of magnitude larger while its threshold is in the
GeV, rather than the MeV range.

Detectors presently under construction have a nominal effective area of
$2\times10^4$~m$^2$. This area depends on the energy of the muon and the
trigger; for the AMANDA detector it varies from 0.6--$3\times10^4$\,m$^2$ for
1--100~TeV muon energy\cite{Tilav}. The DUMAND area is about
$1.2\times10^4$\,m$^2$ at 1~TeV averaged over the lower hemisphere, and grows
roughly logarithmically with energy to an area of over $4\times10^4$\,m$^2$
at 100~TeV\cite{OKA}.

The difference between DUMAND and AMANDA in energy dependence is due to the
difference in predicted optical properties of ice and deep seawater, and the
different size of photomultipliers used in the two experiments. The threshold
is in the 2--10~GeV energy range for AMANDA and is about 10~GeV for DUMAND.
Both detectors also have a large detection volume for showers initiated by
$W$'s produced by electron-antineutrinos on atomic electrons at 6.4 PeV, with
DUMAND reaching $\sim0.2$\,km$^3$.

Relative to a 1~km scale detector, the experiments under construction are
only ``few" percent prototypes. Yet, using natural water or ice as a detection
medium, these neutrino detectors can be deployed at roughly 1\% of the cost
of conventional accelerator-based neutrino detectors which use shielding and
some variety of tracking chambers.  It is thus not hard to believe that the
Cherenkov detectors can be extended to a larger scale at reasonable cost.  The
first generation telescopes consist of roughly 200
optical modules (OM) sensing the Cherenkov light of cosmic muons. The
experimental advantages and challenges are different for each experiment and,
in this sense, they nicely complement one another.  Briefly,

\begin{itemize}
\item  DUMAND is positioned under 4.5~km of ocean water, below most biological
activity and well shielded from cosmic ray muon backgrounds.  One nuisance
of the ocean is the background light resulting from radioactive decays, mostly
K$^{40}$, plus some bioluminescence, yielding an OM
noise rate of  50--100~kHz.  On the other hand, deep ocean water is
fantastically clear, with an attenuation length of order 40~m in the
blue\cite{DUM}.  The deep ocean is stable, quiet, and
completely shielded from electromagnetic interference.  There is also an
unlimited quantity of territory within 30~km of easily accessible and
habitable shore locations.  Yet, the deep ocean is a difficult location for
access and service, not at all like a laboratory experiment.  Detection
equipment must be built to high reliability standards, and the data must be
transmitted to the shore station for processing.  It has required years to
develop the necessary technology and learn to work in an environment foreign
to high-energy physics experimentation, but hopefully that is now accomplished
satisfactorily.

\item AMANDA is operating in deep clear bubble-free ice. The ice provides a
convenient mechanical support for the detector. The immediate advantage is
that all electronics can be positioned at the surface. Only the optical
modules are deployed into the deep ice. Polar ice is a sterile medium with a
concentration of radioactive elements reduced by more than $10^{-4}$ compared
to sea or lake water. The low background results into an improved sensitivity
which allows for the detection of high energy muons with very simple trigger
schemes which are implemented by off-the-shelf electronics. Being positioned
under only 1~km of ice it is operating in a cosmic ray muon background which
is over 100 times larger than DUMAND. The challenge is to reject the
down-going muon background relative to the up-coming neutrino-induced muons by
a factor larger than $10^6$. The group claims to have met this challenge with
an up/down rejection which is at present superior to that of the deep
detectors\cite{Tilav}. The polar environment is difficult as well, with
restricted access and one shot deployment of photomultiplier strings.  The ice
may not be as clear as deep ocean water. It has, however, been shown by three
independent methods that in-situ polar ice has an attenuation length exceeding
20~meters at 800~meters depth\cite{LOW}. A shorter optical attenuation length
in ice compared to seawater may or may not be a limitation in that the AMANDA
8~inch photomultiplier tubes do not see muons far enough to suffer that limit.
 In the ocean, the economic studies indicate that larger detectors are
optimal. The economics of hot water hole drilling (fuel cost limited) dictate
limited depth and diameter of holes. We note that future technology employing
long cylindrical photomultipliers could be a nice match to this problem.

\item NESTOR is similar to DUMAND in being placed in the deep ocean (the
Mediterranean), except for two critical differences.  Half of its optical
modules point up, half down. The angular response of the detector is being
tuned to be much more isotropic than either AMANDA or DUMAND, which will give
it advantages in, for instance, the study of neutrino oscillations. Secondly,
NESTOR will have a higher density of photocathode (in some substantial volume)
than the other detectors, and will be able to make local coincidences on lower
energy events, even perhaps down to the supernova energy range (tens of
MeV)\cite{RES}.

\item BAIKAL shares the shallow depth with AMANDA, and has half its optical
modules pointing up like NESTOR\cite{BAI}.  It is in a lake with 1.4~km
bottom, so it cannot expand downwards and will have to grow horizontally.
Optical backgrounds similar in magnitude to ocean water have been discovered
in Lake Baikal. They vary with season and their nature is not well understood.
 The difficulties in Russia make progress somewhat uncertain, but even so the
Baikal group has deployed an array with 36 Quasar photomultiplier (a
Russian-made 15~inch tube) units in April 1993, and may well count the first
neutrinos in a natural water Cherenkov detector (though with an area similar
to the IMB mine based instrument).

\item Other detectors have been proposed for near surface lakes or ponds
(e.g.\break GRANDE, LENA, NET, PAN and the Blue Lake Project), but at this time
none are in construction\cite{LEA92}.  These detectors all would have the
great advantage of accessibility and ability for dual use as extensive air
shower detectors, but suffer from the $10^{10}$--10$^{11}$ down-to-up ratio of
muons, and face great civil engineering costs (for water systems and light
tight containers). Even if any of these are built it would seem that the costs
would be too large to contemplate a full km scale detector.
\end{itemize}

In summary, there are four major experiments proceeding with construction,
each of which have different strengths and face different challenges. All have
successfully operated small prototypes, all have funding at some level to
proceed, and all four may well be operating with detectors in the few times
$10^4$\,m$^2$ by 1996 or so.  It is thus not too soon to begin to contemplate
the next stage.

For the construction of a 1~km scale detector one can imagine\cite{FHVenice}
any of the above detectors being the basic building block for the ultimate
1~km$^3$ telescope. The present AMANDA design, for example, consists of 9
strings on a 30~meter radius circle with a string at the center (referred to
as a $1+9$ configuration).  Each string contains 20 OMs separated by 12~m.
 Imagine AMANDA ``supermodules'' which are obtained by extending the basic
string length (and module count per string) by a factor of 4.5.  Supermodules
would then consist of $1+9$ strings with, on each string, 90 OMs separated by
12~meters for a length of 1080~meters. A 1~km scale detector then might
consist of a $1+7$ configuration of supermodules, with the 7 supermodules
distributed on a circle of radius 540~meters, and have a total of about 7200
phototubes. Such a detector can be operated in a dual mode:

\begin{center}
\epsfxsize=4in\hspace{0in}\epsffile{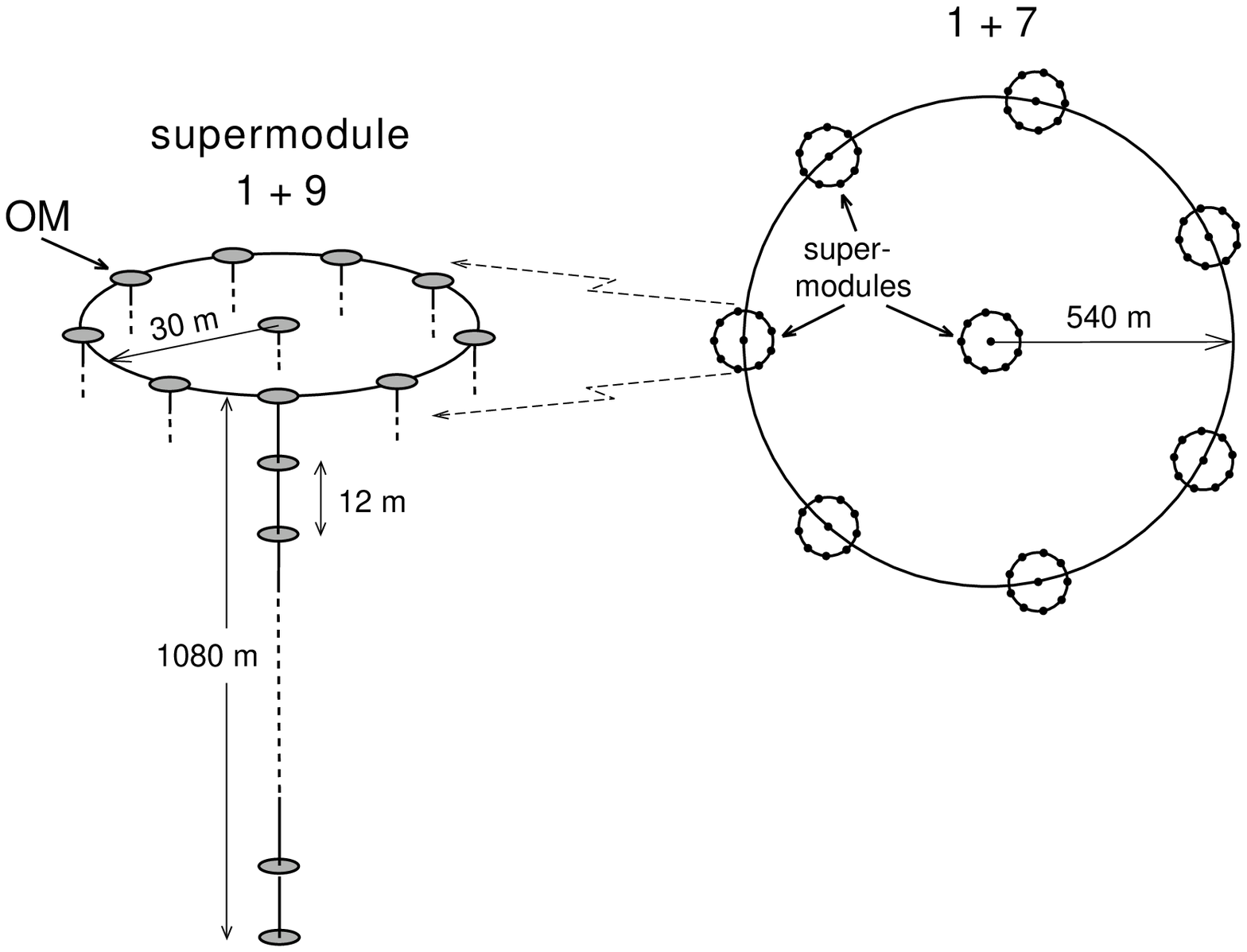}

{\small Fig.~6}
\end{center}

\begin{itemize}
\item it obviously consists of $4.5\times8$ of the presently planned
AMANDA array modules, leading to an effective area
of $\sim0.75$~km$^2$.  Importantly, the characteristics of the detector,
including threshold, are the same as those of the original AMANDA array module.

\item the $1+7$ Supermodule configuration, looked at as a whole, instruments
with
OMs a 1~km$^3$ cylinder with a diameter and height of 1080~m. High-energy
muons will be superbly reconstructed as they can produce triggers in 2 or more
of the modules spaced by large distance.  Reaching more than one supermodule
requires 100~GeV energy to cross 500~m.  For a 1~km deep detector the
threshold for downgoing muons is thus raised from 200 to 300~GeV.  We note
that this is the
energy for which a neutrino telescope has optimal sensitivity to a typical
$E^{-2}$ source (background falls with threshold energy, and until about
1~TeV little signal is lost).
 \end{itemize}

Alternate methods to reach the 1~km scale have been discussed by Learned and
Roberts\cite{ROB}.

How realistic are the construction costs for such a detector?  AMANDA's OMs
cost
roughly \$4K for each photomultiplier tube, pressure vessel, cable and
electronics.  Therefore an array module of 200 OMs costs \$0.8M and the final
detector about \$30M in hardware and data acquisition.  Even if one doubles
the OM costs to account for construction and deployment, the final cost of the
postulated $1+7$ array of supermodules is still below that of Superkamioka
(with $11{,}200 \times 20$~inch photomultiplier tubes in a 40~m diameter by
40~m high
stainless steel tank in a deep mine).
Using deep water and large OMs as in DUMAND~II, (i.e.\ 15~inch instead
of 8~inch AMANDA PMTs), increases the cost per OM by a factor of 2 relative to
AMANDA.  However, the effective muon capture area per optical module is also
increased, by perhaps a factor of ten.  Deployment in deep water involves
additional hardware costs such as junction boxes and cables bringing data to
shore (best estimate is about \$10k/OM amortizing all costs).

Both types of detectors would realize some economy
of large numbers.  In any case, we offer no judgement here on the choice of
basic type of module, array style, nor the choice of experimental location.
Actual performance of the {\it in-situ} detectors will be available in two
years, so a better comparison awaits field experience.   The important point
we do want to stress is that the costs, with these two rather different
approaches we have sketched, are rather close.  No matter which way
you wish to estimate, the cost for a 1~km$^2$ array seems likely to be under
\$100M.  Since this is of the same order as the SUPERKAMIOKANDE detector
presently under construction in Japan such a cost for a non-accelerator
project is certainly not unprecedented nor outrageous by present high energy
physics standards.

\section{Summary and Plea}

The imminent activation of high energy neutrino experiments in Baikal, Greece,
Hawaii and South Pole will provide an ideal testing ground for the design of an
``array module'' and an evaluation of the variations in hardware design, depth
and medium for deployment.  Having effective areas for high energy neutrinos
exceeding existing underground detectors by about two orders of magnitude they
will certainly open new frontiers in cosmic muon and neutrino physics.
Hopefully these experiments will make serendipitous discoveries.  Yet, we must
acknowledge that those instruments now
building for operation by 1995 will probably not be large enough to really
undertake neutrino astronomy.  It will require another step of about two
orders of magnitude to be well into business, and for that it is not too early
for dreams, plans, and studies.

We strongly feel that the construction of the next generation detector should
be a collaborative effort of all the institutions involved in high-energy
neutrino astronomy, a world cooperative effort.  It might even consist of 2
parts, possibly using different techniques, with complementary sky coverage
and independent operation such that the usual checks can be made!

At the 1~km$^2$ size it seems inescapable, based upon present calculations and
simple energetic considerations extrapolating from observations with gamma
rays and lower-energy photons, that such point sources must be seen. At this
size then, one would truly be in the business of astronomy: not just counting
sources but able to observe a multiplicity of sources with enough statistics
to begin extracting information from energy spectra and temporal behaviour,
particularly in comparison with photon observations (e.g.\ in the episodic
behaviour of AGNs, and the timing in binaries).  Indeed the suggestion that
one may be able to ``neutrino ray'' the companion star in a galactic X-ray
binary (such as Her~X-1), could mean that we know more about the density
profile of a distant star than about that of our own sun. It is unlikely that
these observations can be carried out with detectors smaller than 1~km$^2$.
And then there are all the other possibilities\dots

\section*{Acknowledgements}

This research was supported in part by the University of Wisconsin Research
Committee with funds granted by the Wisconsin Alumni Research Foundation, in
part by the U.S.~Department of Energy under Contract No.~DE-AC02-76ER00881 and
in part by the Texas National Research Laboratory Commission under Grant
No.~RGFY93-221.


\end{document}